\documentclass{cjpsuplf}    
\usepackage{graphics}
\usepackage{graphicx}
\usepackage[dvips]{epsfig}
\usepackage{rotating}         
\begin{document}

\title{Heavy Meson Production in NN Collisions with Polarized
  Beam and Target --- A new facility for COSY}

\authori{F.~Rathmann$^1$\footnote{email: \texttt{f.rathmann@fz-juelich.de}}, 
M. D\"uren$^2$, 
P. Jansen$^3$, 
F. Klehr$^3$, 
S. Martin$^1$, \newline
H.O. Meyer$^4$, 
K. Rith$^5$, 
H. Seyfarth$^1$,
E. Steffens$^5$, 
H. Str\"oher$^1$}
\addressi{
$^1$Institut f\"ur Kernphysik, Forschungszentrum J\"ulich, 52425  J\"ulich, Germany\\
$^2$II. Physikalisches Institut, Justus--Liebig--Universit\"at, 35392 Gie{\ss}en, Germany\\
$^3$Zentralabteilung Technologie, Forschungszentrum J\"ulich, 52425  J\"ulich, Germany\\ 
$^4$Department of Physics, Indiana University, Bloomington, Indiana, USA\\
$^5$Physikalisches Institut II, Universit\"at Erlangen--N\"urnberg, 91058 Erlangen, Germany}
\authorii{}     
\addressii{}
\authoriii{}     
\addressiii{}
\headtitle{Study of Heavy Meson Production \ldots}
\headauthor{F. Rathmann et al.}  
\specialhead{First Author et al.: Title of the contribution \ldots}
\evidence{}
\daterec{}    
\suppl{A}  \year{2002}
\setcounter{page}{1}
\maketitle

\begin{abstract}
  The study of near--threshold meson production in $pp$ and $pd$
  collisions involving polarized beams and polarized targets offers
  the rare opportunity to gain insight into  short--range
  features of the nucleon--nucleon interaction.  The Cooler Synchrotron
  COSY at FZ--J\"ulich is a unique environment to perform such
  studies.  Measurements of polarization observables require a
  cylindrically symmetrical detector, capable to measure the momenta
  and the directions of outgoing charged hadrons.  The wide energy
  range of COSY leads to momenta of outgoing protons to be detected in
  a single meson production reaction between 300 and 2500~MeV/c. 
  Scattering angles of protons to be covered extend to about
  $45^{\circ}$ in the laboratory system.  An azimuthal angular
  coverage of the device around 98\% seems technically achievable. 
  The required magnetic spectrometer could consist of a
  superconducting toroid, providing fields around 3~T.
\end{abstract}

\section{Scientific Motivation}
The production of heavy mesons near threshold in $pp$ and $pd$
collisions with stored polarized beams and internal gas targets offers
the rare opportunity to study short--range features of the
nucleon--nucleon interaction, at distances where the nucleons start to
overlap.  The identification of relevant sub--nucleonic degrees of
freedom of the nucleon--nucleon system in the non--perturbative region
of QCD is of great topical relevance to nuclear and particle physics. 
COSY represents a unique environment to perform studies of this kind,
in particular measurements of polarization observables in heavy meson
production reactions, as emphasized recently \cite{iesp}. 
Measurements of those polarization observables are to be considered
which are associated with the spins of the particles in the entrance
channel, since the polarization of particles in a final state in most
cases is difficult to access.

\subsection{Meson production in proton--proton collisions}
When producing a meson close to threshold, the transferred momentum is
large.  In addition, angular--momentum constraints require a head--on
collision (small impact parameter).  From this, it is clear that such
a reaction is sensitive to the NN interaction at distances of less
than 1 fm, while the long--range part is suppressed, thus the reaction
acts as a filter for short--range features of the nucleon--nucleon
interaction.  In hindsight, we can understand why the first precise
measurement of the total cross section in $pp \to pp \pi^0$
\cite{HOM92} resulted in a challenge for theorists.  For the very same
reasons, near--threshold production of mesons heavier than the pion is
a promising field of nuclear research.  In addition, the interaction
of the produced meson with the hadrons in the exit channel can be
studied, and one expects to learn about the coupling of mesons to
nucleons.

The production of a meson in NN collisions close to the production
threshold involves only a few transitions between a given initial and
final angular momentum state.  These transitions can be labeled by the
angular momentum of the NN pair ($L'$) and the angular momentum of the
meson with respect to the nucleon pair ($l'_{meson}$) in the exit
channel.  Close to threshold $l'_{meson}$ and $L'$ can be either 0 or
1.  This scheme results for $pp \to pp\pi^0$ for instance in the
possible combinations $Ss$, $Ps$ and $Pp$ listed in
table~\ref{table1}.  Parity and angular--momentum conservation forbid
$Sp$ final states.  As the incident proton energy is lowered further,
eventually only a single transition into $Ss$ becomes possible. 
\refstepcounter{table} \setcounter{table}{0}
\begin{table}[htb]
    \begin{center}
    \begin{tabular}{l|l|l}
    initial state       & final state                      & class\\
    $^{2S+1}L_J$        & ${(^{2S'+1}L'_{J'},l'_\pi)_J}$    & $L'l'_\pi$\\\hline
    $^3P_0$             & $(^1S_0,s)_0$                    & $Ss$ \\ \hline
    $^1S_0$             & $(^3P_0,s)_0$                    & $Ps$ \\
    $^1D_2$             & $(^3P_2,s)_2$                    &\\ \hline
    $^3P_1$             & $(^3P_0,p)_1$                    & $Pp$ \\
    $^3P_{0,1,2}$       & $(^3P_{1,2},p)_{0,1,2}$          &\\
    $^3F_{2,3}$         & $(^3P_{1,2},p)_{2,3}$             &
    \end{tabular}
    \protect\parbox{12cm}{\caption{\small Contributing amplitudes in
        the process $pp \to pp\pi^0$. The angular--momentum quantum
        numbers in the final state are denoted by a prime.}}
  \end{center}
\label{table1}
\end{table}
These angular--momentum constraints apply also for the case of the
production of heavier pseudo--scalar mesons, e.g. $\eta$ and $\eta'$
($J^P=0^-$), while for the vector mesons, like $\rho$, $\omega$,
$\phi$ ($J^P=1^-$), analogous relations can be worked out.  The total
cross section can be written as an incoherent sum of the cross
sections of the individual final states $\sigma_{tot}=\sigma_{Ss}
+\sigma_{Ps} + \sigma_{Pp}$.  Over a limited energy range close to
threshold, the transition matrix element can be taken as constant.  In
this case, the energy dependence of the reaction is given by phase
space, modified by the final--state interaction and the momentum
dependence of the radial part of the wavefunction.  In $pp \to pp
\pi^0$, the final--state interaction is dominated by the (well--known)
interaction between the two nucleons, but in the production of heavier
mesons cases exist, where the (less well--known) meson--nucleon
interaction plays an important role.  The contribution of the radial
wave functions to the energy dependence differs for different
angular--momentum configurations in the exit channel.  When only the
cross section is measured, the energy dependence of the cross section
is the only handle one has on a decomposition of the reaction in terms
of individual amplitudes.  On the other hand, when polarization
observables are available, a much cleaner and model--independent
amplitude analysis becomes possible.  A polarization analysis of
particles in a final state requires a second scattering, which is
difficult, except in a self--analyzing decay (e.g. $\Lambda^{0} \to
\pi^{-}p$) The polarization observables accessible are those that are
associated with the spins of the two protons in the entrance channel.

On the experimental side at present only the PINTEX group at IUCF
performs studies of this kind in $\vec{p}\vec{p}\to pp \pi^{0}$
\cite{hom98,hom99,hom01}, $\vec{p}\vec{p}\to pn\pi^{+}$
\cite{saha99,daehnick02}, and $\vec{p}\vec{p} \to d\pi^{+}$ reactions
\cite{prz99}, while measurements in $\vec{p}\vec{d} \to t\pi^{+}$,
$\vec{p}\vec{d} \to pd\pi^{0}$, and $ \vec{p}\vec{d} \to {\rm
  ^3He}\pi^{0}$ have not yet been published.  With the apparatus
described in ref.~\cite{rinc00}, it becomes possible to
\textit{experimentally} separate contributions from different angular
momentum states, free of any model \cite{hom99}.  Theoretical
estimates of polarization observables in near threshold pion
production are scarce, and those available still agree not well with
the measurements.  While the agreement for final states with charged
pions is better, in particular the measured spin correlation
parameters for $\vec{p}\vec{p}\to pp \pi^{0}$ are not well reproduced
\cite{hanhart98}.  One is inclined to state that the theoretical
understanding of one of the most fundamental processes in nuclear
physics is still incomplete, and that more insight must come from
measurements of polarization observables.
  
In the case of pion production, measurements with vector-- and
tensor--polarized deuteron and vector--polarized proton targets are
presently carried out at IUCF and will be continued into the near
future.  For the production of heavier mesons, COSY, due to its higher
beam energies and intensities, is in a position to dominate the field
during the years to come, once an appropriate detector system would be
available.  One purpose of such a new experimental facility, recently
proposed for the internal beam of COSY \cite{rath99}, would be a
detailed study of near threshold meson production using a cooled,
stored, polarized proton (or deuteron) beam incident on an internal
polarized hydrogen (or deuterium) storage cell gas target
\cite{rath-rev-prag}.  For $pp\to ppX$ reactions the production of
$\pi^{0}$, $\eta$, $\rho$, $\omega$, $\eta'$ and $\phi$ mesons is
energetically possible at COSY (see table~\ref{table2}).

\refstepcounter{table} \setcounter{table}{1}
\begin{table}[htb]
    \begin{center}
      \begin{tabular}{c|c|c|c}
                  &             &\multicolumn{2}{c}{Threshold}\\
        Meson $X$ & Mass [MeV]  &$P_{lab}~[MeV/c]$  & $E_{lab}$~[MeV]\\\hline
        $\pi^0$   & 135.0      &  776.5             &  279.7 \\
        $\eta$    & 547.5      & 1982.1             & 1254.7 \\
        $\rho$    & 768.5      & 2627.5             & 1851.7 \\
        $\omega$  & 781.9      & 2667.8             & 1889.7 \\
        $\eta'$   & 957.8      & 3208.3             & 2404.4 \\
        $\phi$    & 1019.4     & 3403.9             & 2592.6 
      \end{tabular}
          \protect\parbox{12cm}{\small \caption{Threshold values of
          bombarding proton momenta (and kinetic energies) required to
          produce the listed mesons in the reaction $\vec{p}\vec{p}
          \to ppX$.}}
      \label{table2}
    \end{center}
  \end{table}

\subsection{Meson production in the three--nucleon system}
Na\"{\i}vely, one would think that near--threshold meson production in
$pd$ collisions can be fully understood in terms of a short--ranged
$\rm NN \to NN\pi$ process, with the extra nucleon acting as a
spectator.  Recently, however, pion production in $pd$ collisions has
revealed departures from such a simple picture \cite{rohdjess94}.
These studies are in their infancy and deserve more attention from the
theoretical community.  Since the number of participating amplitudes
in this case, too, is small, even though the transferred momentum is
large, we have a unique environment in which to study the interaction
of more than just two nucleons, e.g.  three--nucleon forces.  It is
likely that eventually the research on three--nucleon forces will
focus on this process.  Again, the role of polarization observables is
important, but practically no such data are available.  In the case of
pion production, measurements with vector-- and tensor polarized
deuterons and polarized protons are carried out at IUCF. For the
production of heavier mesons, COSY again is in a position to dominate
the field in the years to come.

\subsection{Other reactions}
Even though, clearly, the main motivation in designing a new facility
is arising from the need for polarization observables in meson
production, one needs to also mention future possibilities, which
would include the $pd$ breakup reaction \cite{rat-break-prag} as well
as radiative transitions.

Besides what has been presented so far, a truly unique application of a
gas target would be the use of polarized deuterons as an effective
polarized neutron target by detection of low energetic spectator
protons.  The purpose of these measurements would be the generation of
a solid $\vec{p}\vec{n}$ database and, as mentioned above, a detailed
study of meson production in $pd$ collisions.  These studies hinge on
the availability of a spectator detector system, by which energy {\it
and} angle information about the spectator proton have to be measured
along the extended storage cell target \cite{vertex}.

In addition, three body final states involving strange mesons
($K^{\pm,0 }$) and baryons ($\Lambda, \Sigma^{\pm,0}$), for instance
in the reaction $\vec{p}\vec{p} \to p K^+ \Lambda^0$, could be studied
with polarized beam and target.

\section{A new facility for COSY}

Near threshold only a limited angular range in the forward direction
needs to be covered.  However, the momenta of the particles in the
exit channel increase with the threshold of the mesons under study. 
While for pion production near threshold the involved proton momenta
are small and a scintillator detector is sufficient, heavier meson
production experiments with high proton momenta require magnetic
separation\footnote{The range of a proton of 1~GeV kinetic energy,
e.g. in plexiglass (Lucite) is about 2.8~m and the hadronic
interaction probability is $\sim 0.97$ \cite{janni82}.}.  Especially
for experiments involving polarized beams and targets, the azimuthal
angular distribution of the ejectiles depends on the spin orientation
of beam and target polarization.  Thus, measurements of polarization
observables in meson production reactions require a cylindrically
symmetrical detector which is able to measure the momenta and the
directions of outgoing charged hadrons.  It should be pointed out that
for this task a toroidal field configuration is advantageous,
compared to e.g. a solenoid, because residual field components along
the beam axis are absent.  Thus the polarized gas target can be
operated in a weak magnetic guide field.  For the same reason
distortions of the beam closed orbit due to the magnetic field of the
toroid are avoided to a large extent.

\begin{figure}[hbt] 
\begin{center}
\epsfig{file=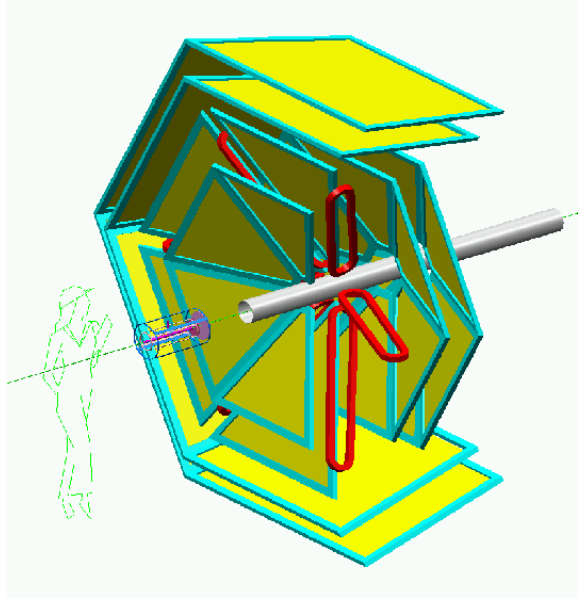,width=7.6cm}
\end{center}
\caption{\footnotesize 3D view of a detector system based on a
toroidal magnetic field configuration.  Six superconducting coils form
a toroid, particle tracking is accomplished by a system of drift and
proportional chambers before and behind the deflecting magnet.  The
polarized storage cell gas target is located in front of the
detector.}
\label{fig:toroid}
\end{figure}

The IUCF pion production experiments have shown that in particular
large values of the kinematical $\eta$ parameter\footnote{$\eta$ is
  defined as the maximum meson momentum divided by the meson mass,
  $\eta=0$ characterizes threshold.} are of interest.  For instance in
$\vec{p}\vec{p} \to pp \pi^0$, only for values of $\eta \sim 0.5
\ldots 1$ higher partial waves begin to play a role.  Large values of
$\eta \sim 1$ correspond typically to proton scattering angles
$\theta_{lab} \sim 30^\circ$.  Consequently, if for the production of
heavy mesons, $\eta$ is the relevant parameter, in order to study the
onset of higher partial waves, the new apparatus should be capable to
accept a large range of scattering angles of the outgoing hadrons up
to about $\theta_{lab}\sim 45^\circ$.  Based on the boundary
conditions outlined above, an axially symmetric toroidal magnetic
spectrometer seems appropriate\footnote{Similar magnet constructions
  of two recently commissioned devices, HADES at GSI \cite{hades} and
  CLAS at Jefferson Lab \cite{clas}, are based on six single helium
  cooled superconducting loops arranged such as to form a toroid.  The
  BLAST setup \cite{blast} employs warm coils.}.  The first concept of
such a device is depicted in Fig.~\ref{fig:toroid}.

\section{Conclusion}
The recently approved upgrade of the beam injection system, once
installed, will supply COSY with high polarized proton and deuteron
beam intensities up to the acceptance limit of the machine.  An
intense polarized light--ion beam stored in COSY, bombarding a purely
polarized internal gas target, surrounded by a toroidal magnetic
spectrometer, capable to detect charged hadrons of high momentum, will
allow us to fully exploit the machines scientific potential.  Such a
device would be ideally suited as one of the {\it next generation}
experiments at COSY, providing a {\it Short Range Laboratory of Meson
  Production}.

\end{document}